\documentclass[a4paper,twoside,brazil,english,prd,amsmath,nofootinbib,superscriptaddress,showpacs,twocolumn]{revtex4-1}
\usepackage{ae,aecompl}
\usepackage[T1]{fontenc}
\usepackage[utf8]{inputenc}
\usepackage[usenames,dvipsnames]{color}
\usepackage{babel}
\usepackage{verbatim}
\usepackage{graphicx}
\usepackage{amsmath}
\usepackage{amssymb}
\usepackage[unicode=true,
 bookmarks=true,bookmarksnumbered=false,bookmarksopen=false,
 breaklinks=false,pdfborder={0 0 1},backref=section,colorlinks=true]
 {hyperref}
\hypersetup{pdftitle={paperMTS-TH},
 pdfauthor={AM and MLeD},
 pdftex}

\makeatletter

\pdfpageheight\paperheight
\pdfpagewidth\paperwidth

\@ifundefined{textcolor}{}
{%
 \definecolor{BLACK}{gray}{0}
 \definecolor{WHITE}{gray}{1}
 \definecolor{RED}{rgb}{1,0,0}
 \definecolor{GREEN}{rgb}{0,1,0}
 \definecolor{BLUE}{rgb}{0,0,1}
 \definecolor{CYAN}{cmyk}{1,0,0,0}
 \definecolor{MAGENTA}{cmyk}{0,1,0,0}
 \definecolor{YELLOW}{cmyk}{0,0,1,0}
 }

\@ifundefined{date}{}{\date{}}
\@ifundefined{definecolor}
 {\usepackage{color}}{}
\usepackage{babel}
\usepackage{indentfirst}\usepackage{array}\usepackage{mathrsfs}
\usepackage{enumerate}\usepackage{nicefrac}\newcommand{\ud}{\ensuremath{\mathrm{d}}}
\newcommand{\Lie}{\ensuremath{\mathcal{L}}}

\usepackage{enumitem}

\makeatother

\begin{document}

\title{A dual null formalism for the collapse of fluids in a cosmological background}

\author{Alan Maciel}

\email{amsilva@if.usp.br}

\affiliation{Instituto de Física, Universidade de São Paulo,\\
 Caixa Postal 66.318, 05315-970, São Paulo, Brazil}

\author{Morgan Le~Delliou}

\email{delliou@ift.unesp.br}

\affiliation{Instituto de Física Teorica, Universidade Estadual de São Paulo (IFT-UNESP),\\
 Rua Dr. Bento Teobaldo Ferraz 271, Bloco 2 - Barra Funda, 01140-070
S\~{ã}o Paulo, SP Brazil}

\altaffiliation{also Centro de Astronomia e Astrofísica da Universidade de Lisboa, Faculdade de Ciências, Ed. C8, Campo Grande, 1769-016 Lisboa, Portugal}




\author{José P. Mimoso}

\email{jpmimoso@fc.ul.pt}

\affiliation{Departamento de F\'{\i}sica,  Faculdade de
Ci\^encias da Universidade de Lisboa, and Instituto de Astrof\'{\i}sica e Ci\^{e}ncias do Espa\c{c}o,
Edif\'{\i}cio C8, Campo Grande,
P-1749-016 Lisbon, Portugal}

\date{submitted: 23/06/15; accepted ..; Published ..}

\begin{abstract}
In this work we revisit the definition of Matter Trapping Surfaces (MTS) introduced in previous investigations and show how it can be expressed in  the so-called dual null formalism
developed for Trapping Horizons (TH). With the aim of unifying both approaches, we construct a 2+2 threading from the 1+3 flow, and thus isolate one prefered
spatial direction, that allows straightforward translation into a
dual nul subbasis, and to deduce the geometric apparatus that follows.
We remain as general as possible, reverting to spherical symmetry
only when needed, and express the MTS conditions in terms of 2-expansion of the flow,
then in purely geometric form of the dual null expansions.
The Raychadhuri equations that describe both MTS and TH are written and interpreted using 
the previously defined gTOV (generalized Tolman-Oppenheimer-Volkov) functional introduced in previous work.
Further using the Misner-Sharp mass and
its previous perfect fluid definition, we relate the spatial 2-expansion
to the fluid pressure, density and acceleration. The Raychaudhuri
equations also allows us to define the MTS dynamic condition with
first order differentials so the MTS conditions are now shown to be all first
order differentials. This unified formalism allows one to realise that the MTS
can only exist in normal regions, and so it can exist only between black
hole horizons and cosmological horizons. Finally we obtain a relation
yielding the sign, on a TH, of the non-vanishing null expansion which
determines the nature of the TH from fluid content, and flow characteristics.
The 2+2 unified formalism here investigated thus proves a powerful tool to reveal, in the
future extensions, more of the very rich and subtle relations between MTS and
TH.
\end{abstract}
\maketitle
\section{Introduction}

The study of cosmological expansion in local self-gravitating structures
is an ancient question in General Relativity, dating back to the works
of McVittie \cite{McVittie:1933zz} and Einstein and Straus \cite{Einstein:1945id,Einstein:1946zz}.
The difference between these two pioneering approaches lies in that
the first adopts a smooth metric with appropriate limits at local
and global scales, while the second introduces a junction between
a local solution and a global one. Since then both approaches have
been explored in the literature \cite{Nolan:1999kk,Faraoni:2007es, Nolan:2005tt, Carrera:2008pi,Ellis:2010fr, Guariento:2012ri,daSilva:2012nh,Mars:2013ooa,daSilva:2015mja} (see also \cite{Hochberg:1998ha,Cattoen:2005dx,Gibbons:2009dr,Clifton:2010fr,Clarkson:2011zq,Sussman:2013qya,Barcelo:2006uw} and references therein for related issues).

One alternative 
approach based on smooth solution was explored in the works
of \cite{Mimoso:2009wj,ledelliou-2011, Mimoso:2013iga,Delliou:2013xra}, where, in spherical symmetry,
the concepts of {\em Matter Trapping Shells} or {\em Surfaces} (MTS hereafter),
also referred to as ``separating shells'', was introduced.
This concept, first discussed there in perfect fluids, appears as
a candidate to divide the scales of global physics from those of local
physics in self-gravitating systems with a cosmological background.
The introduction in this framework of imperfect fluids, with anisotropic
stress and heat flux terms, was later proposed in \cite{Mimoso:2013iga,Delliou:2013xra}.

  This approach relies on the idea that one should search within general relativistic gravity for a transition between the cosmological realm characterized by the expansion, and, say, an astrophysical domain, where manifestly the non-linear collapse of overdensities either produces virialized structures or more extreme collapsed objects. This goal is parallel, and somewhat complementary  to understanding why, on smaller scales, structures and  physics altogether seem to be immune to the overall expansion of the universe. However, it does not necessarily require that the physics on the smaller scales be newtonian, albeit it does not exclude it. Black holes and other extreme objects may very well be the outcome of the divide that we envisage. And we should also refer to other interesting possibilities such as the cracking phenomenon put forward by Herrera and collaborators, which is associated to anisotropic instabilities of compact objects in Astrophysics \cite{herrera-1992,herrera-1997b,herrera-2004,herrera-2010a}  


Indeed, a different aspect of collapsing spacetimes solutions is the formation of dynamical black holes, for which there is a robust formalism first proposed in the seminal paper of Hayward \cite{Hayward:1993mw} and further progress in the domain have been steady since then (see \cite{Hayward:1997jp, Hayward:2008jq, Senovilla:2011fk} and references there in). Dynamical black holes are identified and described by the behaviour of the \emph{trapping horizons} (TH).
It is therefore of great interest to keep this viewpoint in mind and explore a formal approach that might share concepts.

In this work we aim at making 
use of the tools that have been proved useful in the study of dynamical black holes in order to study the properties of MTS's in collapsing spacetimes with a cosmological background. We write the equations that define the MTS and those defining the TH in terms of the same scalars and operators, by finding the relationship between the scalars in the 1+3 formalism --- which have been used in the former literature in the MTS --- and the scalars in the  2+2 formalism --- which is mainly used in the TH literature. We then present both points of view in the common 2+2 formalism. We are led to
obtain a form of the MTS conditions which avoids derivatives of the 2-expansions, and that thus establishes  algebraic equations relating the 2-expansions, the curvature of the 2-spheres ( the $r^{-2}$ term) and the energy momentum tensor. This illustrates how the consideration of the role of matter within the present geometrodynamical formalism may be of importance.

Finally we outline further
progresses in interpretations and then conclude.

\section{Theoretical Framework} \label{sec:recall}

We  first recall the main concepts and conditions defining both
MTS and TH, respectively starting from a 1+3 framework and from a
2+2 framework and proposing some clues about translation of these
in the respective framework of each other's concept.

\subsection{Definition of MTS in 1+3 formalism}

\label{sub:Translating-MTS-to}

\subsubsection{Local definition}

In \cite{Mimoso:2009wj,ledelliou-2011,Mimoso:2013iga,Delliou:2013xra},
a formalism
based on a mixture of the 3+1 (ADM \cite{Arnowitt:1959ah}) metric
decomposition and the 1+3 \cite{Ellis:1971pg,Ellis:1998ct,ellis-1999,ellis-2002}
spacetime decomposition was adopted with the purpose of separating local from global physics. This formalism allows one to split quantities along and orthogonal
to the direction of the fluid flow. Following Lasky and Lun \cite{lasky-2006,Lasky:2006zz,Lasky:2007jd,Lasky:2007ky},
the metric follows the form referred to as generalised Painlevé-Gullstrand
(GPG) \cite{lasky-2006,lasky-2007} 
\begin{equation}
\ud s^{2}=-\alpha^{2}\ud t^{2}+\frac{(\beta\ud t+\ud R)^{2}}{1+E(t,R)}+r(t,R)^{2}\ud\Omega^{2}\,.
\end{equation}

\noindent Here  the functions of the temporal and radial
coordinates $t$ and $R$ that we denote by $\alpha,\beta,E$ and $r$, respectively,
are the lapse and radial component of the shift vector (from the nomenclature
of \cite{Arnowitt:1959ah}), the energy-curvature of the frame-flow-orthogonal
hypersurfaces, and the areal radius. Also $\Omega$ represents the solid
angle. For a perfect fluid, 
the MTS are defined by two equations
\cite{Mimoso:2009wj}, establishing distinct conditions. The first, a
kinematic condition, guarantees the instantaneous conservation of the
spherically symmetric Misner-Sharp mass \cite{misner-1964} interior
to the shell in terms of gauge invariant quantities, namely of the
flow's optical scalars, 
\begin{equation}
\frac{\Theta_{3}}{3}+\sigma=\frac{\Lie_{u}r}{r}=0\,,\label{eq:turnaround}
\end{equation}
 where $u=u^{a}$ is the source fluid flow 4-vector. We have introduced the
new notation $\Theta_{3}$ for its expansion, $\sigma$ denotes its
shear scalar as defined in \cite{lasky-2006,lasky-2007}, $\Lie_u$ denotes the Lie derivative along $u$ and $r$
is the areal radius of the MTS. Since $\Lie_u r = u^a \partial_a r$, and the 1-form $n_a \equiv \partial_a r$ is the gradient of $r$, Eq.~\eqref{eq:turnaround} can be
 interpreted as the locus $\mathfrak{K}$ (a hypertube) where the flow is tangent
to a hypersurface foliated by spheres of constant areal radius ($u^a n_a =0$). In other words, the hypertube defined by Eq.~\eqref{eq:turnaround} is foliated by the spheres where the area of the fluid shell reaches an extremal value.

\begin{figure}
\includegraphics[width=0.9\columnwidth]{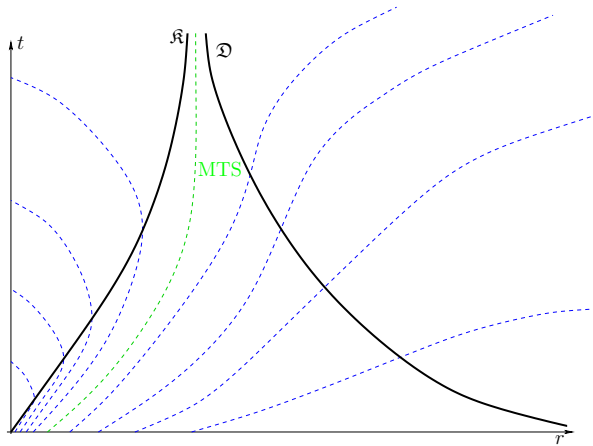}

\protect\caption{\label{fig:KinDynLCDM}A sketch of the worldlines for the second  dust+$\Lambda$
example of Ref.~\cite{Mimoso:2009wj}. The initial condition are typical of cosmology. In this case, the points of $\mathfrak K$ are the areal radius turnaround events while those of $\mathfrak D$ are its zero acceleration events. Both asymptote to the MTS.}
\end{figure}

\noindent The second condition is dynamical, and related to a generalisation
of the hydrostatic equilibrium equation of Tolman-Oppenheimer-Volkoff
(TOV) through a functional that was termed gTOV \cite{Mimoso:2009wj}.
It reduces to 
\begin{align}
-\frac{\text{gTOV}}{r}\equiv \Lie_{u}\left(\frac{\Theta_{3}}{3}+\sigma\right)+\left(\frac{\Theta_{3}}{3}+\sigma\right)^{2} =\frac{\Lie^2_u r}{r}=0,\label{eq:gTOV}
\end{align}
 in terms of gauge invariants from the flow (or the locus $\mathfrak{D}$ where $u^a \partial_a\left(u^b n_b\right)=u^a n_{_2 a}=0$). Eq.~\eqref{eq:gTOV}
reduces to the TOV condition for perfect fluids when Eq.~\eqref{eq:turnaround}
is satisfied on the same shell. In contrast with the use of the TOV
equation in stellar-type solutions, where it holds for all shells
with matter, in the present context it only holds on individual shells where
Eqs.~\eqref{eq:gTOV} and \eqref{eq:turnaround} are satisfied. The
former acts as an equilibrium condition, and is completed with Eq.~\eqref{eq:turnaround}
as a kinematic constraint to keep the shells from escaping equilibrium.
If the two conditions, \eqref{eq:turnaround} and \eqref{eq:gTOV}
are satisfied simultaneously, we have a dividing shell that separates
locally an inner region that can collapse from an outer region that
can expand. Adding initial and boundary conditions of cosmological
expansion, it is argued that such division is global. Such dividing
shell was termed MTS and examples of spherically symmetric solutions
can be found in the references \cite{Mimoso:2009wj,ledelliou-2011,Mimoso:2013iga,Delliou:2013xra}. 
One such example is the simplest spherically symmetric model of dust
with a cosmological constant, using a cosmologically inspired initial
density profile and a Hubble flow to determine the spacetime completely.
As an illustration, we present some flow lines for such model in Fig.~\ref{fig:KinDynLCDM},
illustrating the hypertubes $\mathfrak{K}$ and $\mathfrak{D}$ as
well as the flow line defining the MTS in that case.

\subsubsection{Global definition}

In previous works \cite{Mimoso:2009wj,ledelliou-2011,Mimoso:2013iga,Delliou:2013xra},
the local definition was used to characterise surfaces that would
globally separate expansion from collapse with the crucial assumptions
that some FLRW model would match at spatial infinity and that initial
velocities should all be directed outwards in a Hubble-like fashion,
as expressed by the requirement of cosmological initial and boundary
conditions. \renewcommand{\theenumi}{\alph{enumi}}
\begin{enumerate}
\item \textbf{Cosmological~boundary~conditions:}\label{enu:CosmologicalBoundaryConditions}
A model is declared to have cosmological boundary conditions if its
initial velocities are all directed outwards and it tends to a given
FLRW model at its spatial infinity.
\item \textbf{Local~Matter~Trapping~Surface~Definition:}\label{enu:LocalMatterTrappedSurfaceDefinit}
A 2-surface is locally called Matter Trapping Surface iff it verifies
Eqs.~(\ref{eq:turnaround}) and (\ref{eq:gTOV}) simultaneously.
\end{enumerate}
What set of events would, with those conditions~\ref{enu:LocalMatterTrappedSurfaceDefinit},
constitute a global definition of an MTS was left open. Aesthetical
considerations would drive one towards a continuous 3-volume made of 2-surfaces
over a nonzero mesurable set of flow trajectories, so that the local conditions~\ref{enu:LocalMatterTrappedSurfaceDefinit}
should happen for continuous extended time. However, in examples presented
in these papers, such requirement is not fulfilled. Instead, as seen
in the $\Lambda$ and dust models of \cite{Mimoso:2009wj,ledelliou-2011},
and in the radiation with dust model of \cite{Mimoso:2013iga}, conditions~\ref{enu:LocalMatterTrappedSurfaceDefinit}
were fulfilled only for a 2-surface of events at time infinity, and
the surface defined and identified only thanks to integrable properties
of these models.

In retrospect, this is not too surprising: the two conditions~\ref{enu:LocalMatterTrappedSurfaceDefinit}
describe necessary properties of the MTS in the shape of 3-world~sheets.
However they are not expected to coincide in a nonzero mesurable set.
On the other hand, as in the examples of those previous papers, conditions~\ref{enu:LocalMatterTrappedSurfaceDefinit}
fulfilled at time infinity define the kind of surface, traced back
with their integrable properties, with the properties we wish to promote
as MTS. For the purpose of this paper we therefore propose to use
the following definition:

\begin{enumerate}[resume]
\item \textbf{Asymptotic~Global~Matter~Trapping~Surface~Definition:\label{enu:TentativeGlobalMatterTrappedSurf}}
A Matter Trapping Surface is defined by a surface that evolves along
flow lines and verifies condition~\ref{enu:LocalMatterTrappedSurfaceDefinit}
at time infinity.
\end{enumerate}
\subsection{Trapping Horizons in Dual Null formalism} \label{ssec:recall-TH}

 A very useful notion for the study of dynamical spacetimes and characterization of black holes in such setups was proposed by Hayward in \cite{Hayward:1993mw}, where the \emph{Trapping Horizons}(TH) where first defined. 
 
 The main idea is that, if the spacetime can be foliated in codimension-two spacelike surfaces $\Sigma_{s,t}$ where $s$ and $t$ are the parameters that identify each surface, then we can describe the Lorentzian two-dimensional submanifold orthogonal to the $\Sigma$'s in terms of dual null coordinates. Therefore, each $\Sigma$ surface can be understood as a congruence of null curves, in two independent null directions.

 We can then classify the 2-surfaces according with the expansion of the null curves across the surface, which we'll call here the 2-expansions, to distinguish from the expansion $\Theta_3$ of the flow vector across its orthogonal three dimensional space. The 2-expansions of the congruences defined by a given surface $\Sigma$ and a vector field $v^a$ are given by
 \begin{gather}
  \Theta_{(v)} \equiv\frac{1}{2} N^{ab} \Lie_{v} N_{ab} |_{\Sigma}=N^{ab} \nabla_a v_b\, ,
 \end{gather}
 where the 2-expansion of the congruence $v$ is noted $\Theta_{(v]}$ and is defined using the projector onto the surface $\Sigma$ orthogonal to the dual null directions, say $k^a$  and $l^a$, generally set up as \begin{gather}
 N_{ab} \equiv g_{ab}+k_{(a} l_{b)}\, .
 \end{gather}In the case of spherical symmetry, we can use the areal radius scalar, that is constant on each sphere, to write the two-expansions of a vector field $v^a$ across the spheres:
\begin{gather}
\Theta_{(v)}= \frac{2}{r} \Lie_v r \,. \label{eq:theta-spherical}
\end{gather}

Choosing two linearly independent null vector fields in the subspace orthogonal to the 2-spheres of symmetry, $k^a$ and $l^a$, we can define three types of spheres according to the value of the null two-expansions across it:
\begin{itemize}
\item If $\Theta_{(k)} \Theta_{(l)} < 0$ on the sphere, it is called a \emph{normal} or \emph{untrapped  surface}. If a region of the spacetime can be foliated in normal surfaces, we call it a \emph{normal region},
\item If $\Theta_{(k)}\Theta_{(l)} > 0$ on the sphere, it is called a \emph{trapped surface}. Depending on the sign of the expansions, a trapped surface can be:
		\begin{itemize}
        \item \emph{future trapped} or simply \emph{trapped}  if $\Theta_{(k)} \, , \Theta_{(l)} < 0$;
        \item \emph{past trapped} or \emph{anti-trapped} if $\Theta_{(k)} \, , \Theta_{(l)} > 0$.
        \end{itemize}
        If a region can be foliated in terms of trapped surfaces we call it a \emph{trapped region}. Trapped regions can also be classified as future or past.
\item If $\Theta_{(k)}\Theta_{(l)} = 0$, then it is called a  \emph{marginal surface}. Without loss of generalily, we assume that $\Theta_{(k)}=0$ on a certain marginal surface. Thus, this surface can still be either
\begin{itemize}
\item \emph{outer}, if $\Lie_l \Theta_{(k)} < 0$;
\item \emph{inner}, if $\Lie_l \Theta_{(k)} > 0$;
\item \emph{degenerated} if $\Lie_l \Theta_{(k)} = 0$;
\end{itemize}
\end{itemize} 

A \emph{trapping horizon} is defined as a Lorentzian 3-hypersurface foliated by marginal surfaces. Trapping horizons can be boundaries between regions of different kind. We can classify trapping horizons as past/future and outer/inner in the same way as marginal surfaces.

In static solutions, TH's coincides with event horizons (EH). But in dynamical solutions, TH and EV correspond to different hypersurfaces in general. Since the EV definition is global and the determination of the EV is only possible after integration of the whole future of the spacetime, in the dynamical case the EV has been replaced by the TH as a definition or characterization of black holes. Black Holes, in the sense of a very compact object that is created as the result of gravitational colapse, are associated with future outer trapping horizons. On the other hand, cosmological horizons are past inner trapping horizons in this formalism. For further detail on Black Holes dynamics and TH theory we refer the reader to \cite{Hayward:1993mw,Hayward:1994bu,Hayward:1997jp,Hochberg:1998ha, Mars:2003ud,Hayward:2008jq,Senovilla:2011fk, Senovilla:2014ika}

\section{Translating 1+3 to 2+2 formalisms }

\subsection{MTS definition adapted to a codimension two foliation}


In the following discussion, we will expand the framework of MTS presented
in Sec.~\ref{sub:Translating-MTS-to} and express its definition
in a codimension-two formalism, \emph{i.e.}, in a formalism where consider the evolution of scalars defined on the leaves of codimension-two foliation, instead of the usual formalism, which defines its scalars in a 3-space orthogonal to the flow, being a codimension-one formalism.

We refer to the codimension-one formalism as the 1+3 formalism, while the codimension-two formalism is referred to as the 2+2 formalism, but depends on the choice of basis in the Lorentzian 2-space orthogonal to the spheres. We can chose a pair timelike/spacelike vector as a basis, or two null linearly independent vectors as our basis, one is easily derived from the other. The passage from 1+3 to a codimension-two formalism is the subtlest step in our work. 

We will first introduce the foliation most natural
to spherical symmetry, then the 1+3 and 2+2 projectors that allow
to separate the flow from its orthogonal hypersurfaces as well as
the symmetry sheet. We will then recall the construction of the flow
optical scalars before using them in an other look at the MTS definition.

\subsubsection{Foliations in spherical symmetry}

\noindent Since, in spherical symmetry, spacetime is naturally foliated
in 2-spheres, that are maximally symmetric, it is fruitful to adopt
a formalism which already explicit such symmetry, that is where the
tangent space to the 2-spheres is separated from their orthogonal
space. Following Clarkson \cite{Clarkson:2002jz,Clarkson:2003af,Clarkson:2007yp},
 we consider
an orthonormal basis $\{u^{a},e^{a},s_{(1)}^{a},s_{(2)}^{a}\}$, where $s_{(1)}^{a}$
and $s_{(2)}^{a}$ form any basis tangent to the 2-spheres, $u^{a}$
is the fluid flow 4-vector and $e^{a}$ is a normalised spacelike
vector such that $u^{a}e_{a}=s^{a}e_{a}=0$, where $s^{a}$ is any
vector tangent to the symmetry 2-spheres. However, this decomposition can apply to any spacetime for which the flow-orthogonal hypersurfaces can be decomposed between a privileged vector field and its codimension-two orthogonal hypersurfaces.

Also, considering its equal suitability to the symmetry of the problem
and usefulness in the study of spherically symmetric solutions as
well as in the formalism of trapping horizons,
  we shall use the  2+2 formalism, by using the basis vector from the above to build a null basis of two future directed and linearly independent vectors, $\{k^a, l^a\}$. Again, the codimension-two hypersurfaces orthogonal to that basis need not, in general, be spherically symmetric.


We restrict ourselves to the case where the sheets
are always spacelike, that is, all their tangent vectors are spacelike.
Here we define the notation and relations between basis vectors and
projectors that shall be of foremost importance for this work.

Starting from the normalised and
orthogonal vector fields $\{u^{a},e^{a}\}$, satisfying:
\begin{gather}
u^{a}u_{a}=-1\,, \label{eq:norma_u}\\
e^{a}e_{a}=1\,,\label{eq:norma_e}\\
u^{a}e_{a}=0\,.\label{eq:orto_u_e}
\end{gather}
 We define a corresponding dual null basis $\{k^{a}\,,l^{a}\,\}$ using
$u^{a}$ and $e^{a}$ as 
\begin{gather}
k^{a}=u^{a}+e^{a}\,,\label{eq:kDef}\\
l^{a}=u^{a}-e^{a}\,,\label{eq:lDef}
\end{gather}
 that satisfies, by construction 
\begin{gather}
k^{a}k_{a}=l^{a}l_{a}=0\,,\\
k^{a}l_{a}=-2 \, .
\end{gather}

\noindent Imposing the constraints \eqref{eq:norma_u}, \eqref{eq:norma_e} and \eqref{eq:orto_u_e} does not defines our basis uniquely,
since we can build a new pair of basis vectors by a local Lorentz boost 
\begin{gather}
u^{\prime a}=\cosh\omega\, u^{a}+\sinh\omega\, e^{a}\nonumber \\
e^{\prime a}=\cosh\omega\, e^{a}+\sinh\omega\, u^{a}\,,
\end{gather}
 where $\omega=\omega(x^{a})$, and get a new orthonormal basis $\{u^{\prime a},e^{\prime a}\}$ satisfying the same constraints.
In the corresponding null basis, the boost translates into 
\begin{gather}
k^{\prime a}=e^{\omega}k^{a}\,,\nonumber \\
l^{\prime a}=e^{-\omega}l^{a}\,.
\end{gather}

\noindent We have then defined two sets of bases for the orthogonal
space of codimension-two foliation, one set being orthonormal, the other
is dual null. We have explicited the constraints and freedoms of redefinition
such basis should enjoy. In this work, we shall choose $u^{a}$ in
the direction of the matter flow, which also sets $e^{a}$ from the
foliation up to a sign and, subsequently, the basis null vectors without
ambiguity.

\subsubsection{Projectors}

\label{sub:ss-projector}

In this section, although we shall remain in the 4-dimensional case,
we will strive to keep the notation independent of dimensionality
as much as possible.

As is customary, for instance in 1+3 \cite{Ellis:1971pg,Ellis:1998ct,ellis-1999,ellis-2002}%
,
the projector onto the subspace orthogonal to $u^{a}$ reads 
\begin{gather}
h_{ab}=g_{ab}+u_{a}u_{b}\,,\label{eq:hDef}
\end{gather}
 and, analogously as in \cite{Clarkson:2002jz,Clarkson:2003af,Clarkson:2007yp} or in \cite{Hayward-1993,Hayward:1993mw,Hayward:1997jp},
the further projector onto the 2-subspace orthogonal to $u^{a}$ and
$e^{a}$ or in $k^a$ and $l^a$ respectively writes

\begin{gather}
N_{ab}=g_{ab}+u_{a}u_{b}-e_{a}e_{b}=g_{ab}+k_{(a}l_{b)}\,. \label{eq:nDef}
\end{gather}

In this framework, we define the notations for contractions of general
rank-2 tensors $A_{ab}$ with respect to these projectors: 
\begin{gather}
A_{h}=h^{ab}A_{ab};\quad A_{N}=N^{ab}A_{ab}\,.
\end{gather}

We are interested in projecting physically meaningful tensorial quantities
(in particular the covariant derivative of the flow) between their
tangential and orthogonal components to the flow. However, we are also interested
in decomposing the orthogonal components into their trace and traceless
parts.\footnote{In the particular case of spherical symmetry, they can both define scalars and the only degree of freedom orthogonal to the flow is along the radial $e^a$, the spacelike normal to the 2-spheres. } . This is achieved with the definition
of the projector on the subspace of symmetric trace free rank 2 tensors on the 3-space, which includes the tracefree operator that leaves unaffected the leaves of the 2-foliation normal to the dual null subspace. We name this latter operator the STFPr (\emph{symmetric trace free projector}):\footnote{In \cite{Clarkson:2002jz,Clarkson:2003af,Clarkson:2007yp}, it can be seen as the traceless combination of the 2-projector with the projector along the spacelike normal to the leaves of the foliation. The interest of this new projector lies in the case of
spherical symmetry, in which we shall restrict, where all trace free symmetric tensors
can be written as $a(x^{a})P_{ab}$, where $a$ is a scalar function. 
In other words this representation is one dimensional.
} 
\begin{gather}
P_{ab}=h_{ab}-h_{c}^{c}e_{a}e_{b}=N_{ab}-N_{c}^{c}e_{a}e_{b}\,.\label{eq:PSTF}
\end{gather}
Following its definition,
we obtain the symmetric trace-free 3-spatial component along that STFPr of a rank 2 tensor 
from the contraction: 
\begin{gather}
a(x^{a})=\frac{P^{ab}A_{ab}}{P_{cd}P^{cd}}\,.\label{eq:PSTFscalar}
\end{gather}
 where, in 4-dimensions, we get 
\begin{equation}
P^{2}\equiv P^{ab}P_{ab}=N_{c}^{c}h_{c}^{c}=6\,.\label{eq:Pdois}
\end{equation}

\noindent Finally, we can show an important identity for the translation
between scalars in 1+3 to 2+2 formalism. We first note $A_{e}\equiv A_{ab}e^{a}e^{b}$
the doubly projected component of $A_{ab}$ along $e^{a}$. Applying
the STFPr to an arbitrary tensor $A_{ab}$, thanks to Eqs. \eqref{eq:PSTF}
and \eqref{eq:PSTFscalar} we get the following identities 
\begin{gather}
P^{ab}A_{ab}=P^{2}a=A_{h}-h_{c}^{c}A_{e}=A_{N}-N_{c}^{c}A_{e}\,.
\end{gather}
 Eliminating $A_{e}$ and solving for $A_{N}$, we obtain 
\begin{equation}
A_{N}=\frac{N_{c}^{c}}{h_{c}^{c}}A_{h}+\frac{P^{2}}{h_{c}^{c}}a\,,
\end{equation}
 and, using \eqref{eq:Pdois}, we finally get 
\begin{equation}
\frac{A_{N}}{N_{c}^{c}}=\frac{A_{h}}{h_{c}^{c}}+a\,.\label{eq:morganId}
\end{equation}

\subsubsection{Flow scalars}

\noindent \label{sub:ss-1+3}

With the aim to study MTS's, we need to investigate the geometric properties
of the flow $u^{a}$, as described by the independent components of
its covariant derivative $\nabla_{a}u_{b}$. Our goal is to define physically meaningful scalars adapted to our formalism.

Specializing in the 4-dimensional case, in the 1+3 formalism, the flow is decomposed as

\begin{gather}
\nabla_{a}u_{b}=-\dot{u}_{b}u_{a}+\frac{\Theta_{3}}{3}h_{ab}+\sigma_{ ab }+\omega_{ab}
\end{gather}
 where $\Theta_{3}\equiv h^{ab}\nabla_{a}u_{b}$ can be recognised
as the volume expansion scalar of the fluid, that we shall also call as the 3-expansion, to distinguish it from the 2-expansion, defined
below; $\sigma_{ab}$ is the shear tensor and $\omega_{ab}$ is its vorticity,
corresponding to its projected antisymmetric component.

In spherical symmetry, there is no vorticity in finite-streamed flows,
thus all symmetric trace-free tensors living on the 3-space orthogonal
to the flow are proportional to $P_{ab}$, which implies $\sigma_{ab}=\sigma(x^{a})P_{ab}$
and we shall redefine the shear scalar as this $\sigma(x^{a})$ from
$\sigma(x^{i})\equiv\frac{P^{ab}\nabla_{a}u_{b}}{P_{cd}P^{cd}}=\frac{1}{6}P^{ab}\nabla_{a}u_{b}$
(recall the classic definition \cite{Ellis:1971pg,Ellis:1998ct,ellis-1999,ellis-2002} noted here
$\sigma_{E}\equiv\sqrt{\frac{1}{2}\sigma^{ab}\sigma_{ab}}=\sqrt{3}\sigma$).

However, comparing with the results of Sec.~\ref{sub:ss-projector}
with the specialisation $A_{ab}=\nabla_{a}u_{b}$, we obtain 
\begin{gather}
A_{h}=\Theta_{3}\,,\quad a=\sigma\label{eq:teta3&sigma}
\end{gather}

The next step is to write the optical scalars of the flow in the codimension-two formalism. From now on we will explicitly suppose spherical symmetry.


The orthogonal subspace is generated by the basis $\{u^{a},e^{a}\}$.
We can define 2-expansions,
describing the relative variation to the sphere's area (2-volume or
surface) following an infinitesimal displacement along each congruence
orthogonal to the sphere (i.e. $u^{a}$ and $e^{a}$). As the spheres
have codimension-two, the
2-expansions of the basis vector are of natural interest: 
\begin{gather}
\Theta_{(u)}=N^{ab}\nabla_{a}u_{b}\,,\nonumber \\
\Theta_{(e)}=N^{ab}\nabla_{a}e_{b}\,.
\end{gather}

We can then construct the mean curvature vector as

\begin{gather}
\mathcal{K}^a\equiv -\Theta_{(u)}u^{a}+\Theta_{(e)}e^{a} = \frac{2}{r} \nabla^a r\,,\label{eq:mean-curvature}
\end{gather}
 such as, for any $v^{b}$ in the orthogonal subspace $(u,e)$, we
have the 2-expansion along $v^{b}$ defined as $\Theta_{(v)}=v^{b}\mathcal{K}_{b}$.

In particular, for the 2+2 basis vectors, we find

\begin{gather}
\Theta_{(k)}=k^{a}\mathcal{K}_{a}=\Theta_{(u)}+\Theta_{(e)}\,,\nonumber \\
\Theta_{(l)}=l^{a}\mathcal{K}_{a}=\Theta_{(u)}-\Theta_{(e)}\,.\label{eq:nullexpansions}
\end{gather}

\noindent In spherical symmetry there is neither 2-shear nor 2-vorticity
as the only non-vanishing component of the flow derivatives 
is proportional to the induced metric on the 2-spheres $N_{ab}$, a consequence of the maximal symmetry of the spheres. In other words the only
non-vanishing component of the flow derivatives is the trace
component (for things living on the 2-spheres) and thus its traceless
parts are vanishing.

\noindent Using the definitions of Sec.~\ref{sub:ss-projector},
and setting $A_{ab}=\nabla_{a}u_{b}$, we realise 
\begin{gather}
A_{N}=\Theta_{(u)}.\label{eq:tetaU}
\end{gather}

Hence, using the identity \eqref{eq:morganId} and Eqs.~\eqref{eq:teta3&sigma}
and \eqref{eq:tetaU}, we obtain 
\begin{gather}
\frac{\Theta_{(u)}}{2}=\left(\frac{\Theta_{3}}{3}+\sigma\right)\,.\label{expansionId}
\end{gather}

The above result is the fundamental relation we need to bring MTS's and TH's to a unified formalism.

\subsubsection{Defining Matter Trapping Surfaces in 2+2}

In the 1+3 language, MTSs are characterized by the system of Eqs.~\eqref{eq:turnaround}
and \eqref{eq:gTOV}. From Eq.~\eqref{expansionId}, we can express
them in terms of 2-expansions as 
\begin{gather}
\Theta_{(u)}=0\,,\label{eq:turninpoint2d}\\
 \Lie_{u}\Theta_{(u)}+\frac{\Theta_{(u)}^{2}}{2}=0\,.\label{eq:gtov2d}
\end{gather}

This is a surprisingly simple form for the conditions, that suggests
that the 2+2 formalism probably best captures the symmetries of
the problem, only requiring derivatives in the flow direction $u^{a}$.
Furthermore, we wish to express these equations in a dual null 2+2
basis such that we can use the dual null formalism usual in black
hole dynamics and TH's for MTS's. Equations~\eqref{eq:turninpoint2d}
and \eqref{eq:gtov2d} in terms of $k^{a}$ and $l^{a}$ expansions
yield 
\begin{gather}
\Theta_{(k)}+\Theta_{(l)}=0\,,\label{eq:turningpoint2null}\\
\Lie_{k}\Theta_{(k)}+\Lie_{l}\Theta_{(l)}+\Lie_{k}\Theta_{(l)}+\Lie_{l}\Theta_{(k)}+\frac{\left(\Theta_{(k)}+\Theta_{(l)}\right)^{2}}{2}=0\,,\label{eq:gtov2null}
\end{gather}
 the equations of MTS in the 2+2 formalism. Moreover, we point out that the left hand side of Eq. \eqref{eq:gtov2null} is equivalent to $-8\, \frac{\text{gTOV}}{r}$.

\subsection{Relation to Matter Sources\label{sub:Relation-to-the}}

The 2+2 projection of EFE can be decomposed into scalar equations
along the null vector $k^{a}$, 
\begin{gather}
\Lie_{k}\Theta_{(k)}-\nu_{(k)}\Theta_{(k)}+\frac{\Theta_{(k)}^{2}}{2}=-8\pi T_{ab}k^{a}k^{b}\,,\label{eq:RayKK}
\end{gather}
with the normalised cross projection of its covariant derivative
\begin{gather}
\nu_{(k)}=\frac{1}{k^{c}l_{c}}k^{a}l^{b}\nabla_{a}k_{b}\,,\label{eq:nuk}
\end{gather}
along the null vector $l^{a}$,
\begin{gather}
\Lie_{l}\Theta_{(l)}-\nu_{(l)}\Theta_{(l)}+\frac{\Theta_{(l)}^{2}}{2}=-8\pi T_{ab}l^{a}l^{b}\,,\label{eq:RayLL}
\end{gather}
with the normalised cross projection of its covariant derivative,
\begin{gather}
\nu_{(l)}=\frac{1}{k^{c}l_{c}}l^{a}k^{b}\nabla_{a}l_{b}\,,\label{eq:nul}
\end{gather}
the latter two equations being usually referred to as the null Raychaudhuri
equations, and the mixed projection of the symmetric EFE that yields
\begin{multline}
\frac{1}{2}\left(\Lie_{k}\Theta_{(l)}+\Lie_{l}\Theta_{(k)}+\Theta_{(l)}\gamma_{l}+\Theta_{(k)}\gamma_{(k)}\right)+\Theta_{(k)}\Theta_{(l)}\\
-\frac{g_{ab}k^{a}l^{b}}{r^{2}}=8\pi T_{ab}l^{a}k^{b}\,,\label{eq:RayKL}
\end{multline}
where the linear coefficients of the expansions are related to the
dual null commutators, their Lie derivatives (recall $[k,l]^{a}=\Lie_{k} l^a$),
\begin{gather}
\gamma_{(k)}=\frac{1}{k_{c}l^{c}}l_{a}[k,l]^{a}\,, \label{eq:gammak}\\
\gamma_{(l)}=\frac{1}{k_{c}l^{c}}k_{a}[l,k]^{a}\label{eq:gammal}\,.
\end{gather}

We can relate the coefficients \eqref{eq:nuk}, \eqref{eq:nul}, \eqref{eq:gammak} and \eqref{eq:gammal} in useful a way. Let's consider the difference $\nu_{(k)} - \gamma_{(k)}$:
\begin{gather}
\nu_{(k)}- \gamma{(k)} = \frac{1}{k^c l_c} \left[ k^al^b  \nabla_a k_c - l_a k^b \nabla_b l^a + l_a l^b  \nabla_b k^a \right] = \nonumber\\
\frac{1}{k^c l_c}\left[ \left(k^a l^b + l^a l^b \right) \nabla_a k_b \right] = \frac{1}{k_c l^c}\left(2u^a l^b \nabla_a k_b \right) = \nonumber\\
- \nu_{(l)} + \gamma_{(l)} \, , \label{eq:nu-lambda}
\end{gather}
\noindent
where we used the fact that $\nabla_a (k^b l_b) =0$ by construction, and the fact that 
the expressions are all symmetric in $k^a$ and $l^a$. Writing Eq. \eqref{eq:nu-lambda} in terms of the space and time orthonormal vector basis, we obtain:
\begin{gather}
\nu_{(k)}-\gamma_{(k)}=\frac{2}{k^{c}l_{c}}u^{a}\left(u^{b}\nabla_{a}e_{b}-e^{b}\nabla_{a}u_{b}\right)=\nonumber \\
\frac{2u^{a}}{k^{c}l_{c}}\left(-2e^{b}\nabla_{a}u_{b}\right)=-\frac{4}{k^{c}l_{c}}e^{b}\dot{u}_{b}=2\mathcal{A}\,,
\end{gather}
\noindent where $\dot{u}^b \equiv u^a \nabla_a u^b$ and we have used the fact that $\nabla_a u^b e_b =0$, by construction, and the fact that we have, in spherical symmetry,
\begin{align}
\dot{u}^{b} & =e^{b}\left(e^{c}\dot{u}_{c}\right)\equiv\mathcal{A}e^{b},\\
\textrm{so }\left|\mathcal{A}\right| & =\sqrt{\dot{u}^{a}\dot{u}_{a}}\,.
\end{align}
Summing all Raychaudhuri equations, Eqs.~(\ref{eq:RayKK}), (\ref{eq:RayLL})
and twice (\ref{eq:RayKL}), one can isolate the Dynamical condition
(\ref{eq:gtov2null}) and the Kinematic condition (\ref{eq:turningpoint2null})
of the MTS showing the MTS to remain with no dynamics, as expressed
in the TH 2+2 formalism. The summed Raychaudhuri equations can be written thus
\begin{multline*}
\left[\Lie_{k}\Theta_{(k)}+\Lie_{l}\Theta_{(l)}+\Lie_{k}\Theta_{(l)}+\Lie_{l}\Theta_{(k)}+\frac{\left(\Theta_{(k)}+\Theta_{(l)}\right)^{2}}{2}\right]\\
+\Theta_{(l)}\left(\gamma_{l}-\nu_{(l)}\right)+\Theta_{(k)}\left(\gamma_{(k)}-\nu_{(k)}\right)+\Theta_{(k)}\Theta_{(l)}-2\frac{k^{a}l_{a}}{r^{2}}\\
=-8\pi T_{ab}\left[k^{a}k^{b}+l^{a}l^{b}-2l^{(a}k^{b)}\right]
\end{multline*}
\begin{multline}
\hspace{-.5cm}\Leftrightarrow\left[\Lie_{k}\Theta_{(k)}+\Lie_{l}\Theta_{(l)}+\Lie_{k}\Theta_{(l)}+\Lie_{l}\Theta_{(k)}+\frac{\left(\Theta_{(k)}+\Theta_{(l)}\right)^{2}}{2}\right]\\
+\frac{4}{r^{2}}+8\pi T_{ab}4e^{a}e^{b}
=-\Theta_{(k)}\Theta_{(l)}+2\mathcal{A}\left(\Theta_{(k)}-\Theta_{(l)}\right)\\
=\Theta_{(k)}^{2}+4\mathcal{A}\Theta_{(k)}-\left[2\mathcal{A}+\Theta_{(k)}\right]\left(\Theta_{(k)}+\Theta_{(l)}\right)\,,\label{eq:Dcondition}
\end{multline}
the latter equation using the definitions (\ref{eq:kDef}), (\ref{eq:lDef}),
(\ref{eq:norma_u}), (\ref{eq:norma_e}) and (\ref{eq:orto_u_e})
in the curvature and source terms. Note that the source  term corresponds, in spherical symmetry, to the isotropic pressure $P\equiv T_{ab}e^{a}e^{b}$, while the spherical curvature has a definite sign. The left hand sign term between square brackets presents the vanishing part of the dynamics condition \eqref{eq:gtov2null}  $\mathfrak{D}$, while the last parenthesis presents the vanishing part of the kinematic condition \eqref{eq:turningpoint2null} $\mathfrak{K}$. Note also that the symmetry between  $k$  and $l$  has to be broken in the writing of Eq.~\eqref{eq:Dcondition} where we isolate $\Theta_{(k)}$ but that this symmetry remains underlying. This combination of the EFE can be used to propose an alternative  gauge invariant 2+2 definition for the gTOV functional in \eqref{eq:gTOV}  as
\begin{multline}
-\frac{ \, \text{gTOV}}{r}\equiv-\frac{1}{2\,r^{2}}-4\pi T_{ab}e^{a}e^{b}-\frac{\Theta_{(k)}\Theta_{(l)}}{8}+ \\
\frac{\mathcal{A}}{4}\left(\Theta_{(k)}-\Theta_{(l)}\right)\,.\label{eq:gTOV2+2}
\end{multline}

At this point it is worthwhile to remind the definition of the Misner-Sharp mass $M$ as \cite{misner-1964, Hayward:1994bu}
\begin{equation}
g^{ab}\partial_a r \partial_b r = 1 - \frac{2M}{r} \, , 
\end{equation}
\noindent
which, considering Eq. \eqref{eq:mean-curvature} and Eq. \eqref{eq:nullexpansions} , allow us to relate the Misner-Sharp mass with the null 2-expansions:
\begin{equation}
\mathcal{K}^a \mathcal{K}_a = -\Theta_k \Theta_l = \frac{4}{r^2}\left(1 - \frac{2M}{r}\right)\, \,.
\end{equation}

We can now point out that the reader can recognise the perfect fluid gTOV of \cite{Mimoso:2009wj},
\begin{equation}
\text{gTOV} = \left[ \frac{1+E}{\rho + P} P' + 4 \pi P r + \frac{M}{r^2}\right]\, ,\label{eq:gTOVoriginal}
\end{equation}
\noindent
where $\rho$  is the energy density of the fluid defined as  $\rho\equiv T^{ab}u_a u_b$ and the $^\prime $ stands for radial derivation, in  Eq.~\eqref{eq:gTOV2+2}. The right hand side of Eq.~\eqref{eq:gTOV2+2} displays the pressure term in its second term, the mass term in its first and third terms leaving the pressure gradient in Eq.~\eqref{eq:gTOVoriginal} for the last term, that can be identified as
\begin{gather}
\frac{\mathcal{A}}{2}\Theta_{(e)}=-\frac{1}{r}\, \times \frac{1+E}{\rho+P}P^{\prime}\,.\end{gather}

\section{Discussion and Conclusions}

The form of the MTS conditions given in Eqs. \eqref{eq:Dcondition} and \eqref{eq:D+Kcondition} has the advantage that the derivatives of the 2-expansions are eliminated, and what remains are algebraic equations relating the 2-expansions, the curvature of the 2-spheres ( the $r^{-2}$ term) and the energy momentum tensor. In terms of coordinates, we are left with a system of first order nonlinear differential equations, which means that we greatly simplify the computation of the $\mathfrak{K}$, $\mathfrak{D}$ and MTS surfaces in a given solution or initial condition.

If we restrict ourselves to the local MTS definition, condition~\ref{enu:LocalMatterTrappedSurfaceDefinit} of Sec.~\ref{sec:recall} , where \eqref{eq:gtov2null} and \eqref{eq:turningpoint2null} apply together,  the dynamical evolution Eq.~\eqref{eq:Dcondition} becomes
(recall then $\Theta_{(l)}=-\Theta_{(k)}$)
\begin{align}
\Theta_{(k)}^{2}+4\mathcal{A}\Theta_{(k)}= & 4\,\left(\frac{1}{r^{2}}+8\pi T_{ab}\,e^{a}e^{b}\right)\,.\label{eq:D+Kcondition}
\end{align}
In this case the MTS equation reduces to just one first order equation, as a kinematic condition.

\noindent In addition, the 2+2 formalism renders explicit that the condition~\eqref{eq:turningpoint2null}
can only be fulfilled in the so-called \emph{normal regions}
of spacetime, which are regions foliated in normal surfaces,
 or on \emph{extremal
horizons}, where 
\begin{gather}
\Theta_{(k)}=\Theta_{(l)}=0\,.
\end{gather}
 This property makes sense intuitively since trapped regions are related 
 to collapse (future trapped) and to cosmological expansion (past trapped).
Furthermore any MTS cannot belong to either regions. Extremal horizons
can be understood as situations when two trapping horizons coincide. In this case, the extremal horizon itself gives a separation
between collapse and expansion. In other words, the MTS has to be contained between interior THs and an exterior TH, i.e. between black holes' and cosmological horizons, as expected.
\noindent

On the other hand, we can use the tools we developed to relate the TH's to the MTS quantities, in the same framework. In fact for a TH characterised by $\Theta_{(k)}=0$, and using the version \eqref{eq:gtov2null} of \eqref{eq:gTOV}, the dynamics simplifies into
\begin{gather}
\mathcal{A}\Theta_{(l)}=4 \, \frac{\text{gTOV}}{r}-\frac{2}{r^{2}}-16\pi P\,, \label{eq:theta-gtov}
\end{gather}
in addition to the relevant evolution equations, Eqs.~(\ref{eq:RayLL}), in that case. Moreover, Eq. \eqref{eq:theta-gtov} is interesting, because it shows that the gTOV functional appears as a balance term against the curvature term, and the pressure term to give the sign of $\Theta_{(l)}$, once the sign of the acceleration is known. We are provided with a relationship between two quantities that convey a notion of contraction or expansion, the gTOV functional and the null 2-expansion $\Theta_{(l)}$ at a TH.

We conclude emphasizing that this formalism suggests the possibility to study more subtle
relations between the trapping horizons and MTSs to describe the relevant
quantities on an equal footing. It can also serve as a basis for an
MTS definition independent of spherical symmetry, in the style of
the formalism developed to study dynamical black holes proposed in
\cite{Hayward:1993mw}.  Our present results, which reveal the important role of the characterization of matter within the present formalism, find a follow up in the further investigation of specific matter contents. The latter was beyond the bare scope of the present work, and  will be addressed in future work to deepen our understanding of the dynamical implications of the interplay between matter and geometry.

\noindent 
\begin{acknowledgments}
We thank C. Molina and D. C. Guariento for helpful discussions.
A. M. was supported
by FAPESP Grant No. 2013/06126-6.
The work of M.Le~D. has been supported by PNPD/CAPES20132029. M.Le~D. also wishes to acknowledge IFT/UNESP.
JPM acknowledges the  financial support of the Funda\c{c}\~{a}o para a Ci\^{e}ncia e Tecnologia through the project EXPL/FIS-AST/1608/2013 and  the research grant UID/FIS/04434/2013.
\end{acknowledgments}
\appendix

\section{Derivation of covariant focusing equations}

The form of the Raychaudhuri equation \eqref{eq:RayKK} and \eqref{eq:RayLL} are obtained from the usual form found, for example, in \cite{hawking},
\begin{gather}
 \Lie_K \Theta_{(K)} = - \frac{\Theta_{(K)}^2}{2} -\sigma_{(K) \, ab} \sigma_{(K)}^{ab}  +  \omega_{(K)\, ab} \omega_{(K)}^{ab} - R_{ab}K^a K^b \, ,\label{eq:Ray-hawking}
\end{gather}
\noindent
 by imposing spherical symmetry, which eliminates the 2-shear and 2-vorticity terms, and relaxing the affinity condition on $K^a$, $K^a \nabla_a K^b = 0$. Considering a colinear but non-affine null vector field $k^a \equiv e^\Gamma K^a$, and replacing it by $K^a$ in Eq. \eqref{eq:Ray-hawking}, we obtain
\begin{gather}
 \Lie_k \Theta_{(k)} - \nu_{(k)} \Theta_{(k)} + \frac{\Theta_{(k)}^2}{2} +R_{ab}k^ak^b = 0 \, ,
\end{gather}
\noindent
with $\nu_{(k)} = k^a \partial_a \Gamma$. Replacing $K^a$ by $k^a$ in the affinity condition $K^a \nabla_a K^b =0$, we obtain
\begin{gather}
 k^a \nabla_a k^b = k^a \partial_a \Gamma k^b = \nu_{(k)} k^b \, .
\end{gather}

We can use the same reasoning on the $l^a$ Raychaudhuri equation in order to obtain Eqs.~\eqref{eq:nuk} and \eqref{eq:nul}.

The mixed $kl$ projection of EFE can also be obtained starting by its expression in a dual null coordinate form obtained, for example, in Ref. \cite{MartinMoruno:2009iu}:
\begin{gather}
\frac{1}{2} \left( \Lie_{-} \Theta_{(+)} + \Lie_{+} \Theta_{(-)}\right) = - \Theta_{(+)} \Theta_{(-)} + \frac{g_{+-}}{r^2} + 8 \pi T_{+-} \, , \label{eq:cross-comuta}
\end{gather}
\noindent
where the spherically symmetric line element takes the form
\begin{gather}
 \ud s^2 = 2g_{+-} \ud u_+ \ud u_- + r^2 (u_+, u_-) \ud \Omega^2 \, .
\end{gather}

This equation is valid for a null set of coordinates $\{u_+, u_-\}$, that satisfies $\left[ \partial_+, \partial_- \right] = 0$. A general dual null basis $\{ k^a , l^a \}$ does not commute and does not correspond to the derivatives along any coordinate. Using the same reasoning as earlier, we suppose that the basis are related in the following way:
\begin{subequations}
\begin{gather}
 k^a = e^{\eta} (\partial_+)^a \,,\label{eq:eta}\\
  l^a = e^{\psi} (\partial_-)^a \, .\label{eq:psi}
 \end{gather}
\end{subequations}

Replacing Eqs.~\eqref{eq:eta} and \eqref{eq:psi} in Eq.~\eqref{eq:cross-comuta}, we obtain
\begin{gather}
 \frac{1}{2} \left( \Lie_k \Theta_{(l)}+ \Lie_l \Theta_{(k)} \right)= - \Theta_{(k)} \Theta_{(l)} + \frac{g_{ab}k^al^b}{r^2} + 8\pi T_{ab} l^a k^b + \nonumber\\
 \frac{1}{2}\left(\Theta_{(l)}k^c \partial_c \psi  +  \Theta_{(k)}l^c \partial_c \eta \right) \, . \label{cross-corrected}
\end{gather}

The last step is to characterize in a covariant manner the functions $k^c\partial_c \psi$ e $q^c\partial_c \eta$ in Eq.~\eqref{cross-corrected}. Using the commutators:
\begin{gather}
 0 = [\partial_+ , \partial_-] = [e^{-\eta}k^a, e^{-\psi}l^a] = \nonumber\\
 = e^{-\eta - \psi} \left\{ [k, l]^b - (k^a \partial_a \psi)l^b +(l^a \partial_a \eta)k^b \right\} \, .
\end{gather}
Finally, we obtain
\begin{gather}
 k^c \partial_c \psi = \frac{ k_a  }{k^c l_c}[k, l]^a \, ,\\
 l^c \partial_c \eta = \frac{ l_a  }{k^c l_c}[l, k]^a \, .
\end{gather}

\selectlanguage{brazil}%
\bibliographystyle{unsrt}
\bibliography{shortnames,referencias}
 \selectlanguage{english}%

\end{document}